# Evaluating virtual laboratory platforms for supporting online information security courses

**Monther Aldwairi**†‡

Jordan University of Science & Technology, Irbid, Jordan†
Zayed University, Abu Dhabi, UAE‡

ABSTRACT: Distance education had existed for a long time, then it has undergone a renaissance with the advent of computers and the Internet. Distance education relied on physically delivered material and assessments to students, who work offline at home. More recently, online learning or e-learning introduced virtual classrooms, assessments, online tests and transformed the classroom an into interactive online classroom. Despite the large number of online degrees offered, face-to-face remained the dominant mode and e-learning was just used to complement the classroom. The Covid-19 pandemic continues to impact higher education, and online learning is a forgone conclusion. However, the digital divide hindered the disadvantaged schools and students' efforts to transition to online learning. As the pandemic continues to change the education landscape, many challenges arise and prevent student from realising the full potential of e-learning. One of those is the access to physical labs in science, engineering, and computer science programs. This study evaluates practical solutions for virtual labs to be used in teaching information security and ethical hacking. The course ran over five semesters, and 164 students were surveyed. The survey measured perceptions, enjoyment, experiences and attitudes towards virtual labs, and the results were supporting adoption and acceptance of virtual labs.

INTRODUCTION

Distance education is not a new idea and historians date it back to education through mail correspondence in the 1700s. In the late 1800s and throughout the 20th century Radio then Television contributed to the popularity of distance education [1]. Up until the 70's of the past century, getting degrees from home, without having to go to campus, was limited to videotaped lectures that are not interactive and expensive to produce. Advancements in telecommunications in the 1980's revolutionised distance learning by enabling interactive virtual classrooms [2]. However, with the advent of computers and the Internet in the 1990s, we started to see more advanced video conferencing, high speed communications and instant messaging technologies that paved the road for online learning, also referred to as e-learning. Learning management systems (LMS) with integrated video conferencing, forums, Wikis, content management, testing and grading systems make it more affordable to offer interactive online education and degrees. Even though online degrees have gained popularity, but they were looked down upon when compared to the traditional face-to-face degrees [3].

Fast track to the current Covid-19 pandemic making e-learning a reality and driving leaders to theorise that higher education will no longer be the same as it used to be before the pandemic. Business models are being reshaped, many educational institutions are undergoing transformations and traditional degrees are being restructured. These discussions have long existed before the pandemic, which however, forced most to speed up the transition. Rather than reacting to a stressing need, most saw Covid-19 as an opportunity to redefine education to meet the needs of both learners and employers [4].

Professors and instructors who long have been the center of the classroom, are now no longer holding the doors open to knowledge and wisdom. As universities offer free and paid Massive Online Open Courses (MOOC) and many platforms offer online tutorials, courses and degrees, the knowledge is out there and available to all [5]. Professors must transition to the facilitator's role, where student learn and solve problems before coming to the classroom, and then discuss these acquired knowledge and skills together under the guidance of the professor. At the same time self-directed learning (SDL) is becoming indispensable skill especially after graduation, where learners are in full control of finding learning resources and developing their own learning plans [6]. Student centric and customized courses and curricula are another agile response to the shifting requirements of students and employers. General education is giving way to multidisciplinary and interdisciplinary education, where students gain expertise in various disciplines using collaborative and practice-heavy teamwork [7].

Higher education institutions are largely unoriginal, with cookie-cutter approach used to create new institutions and programs without much innovation. Universities must differentiate themselves and seek new programs and areas to set themselves apart from other institutions. Governments, students, and employers alike are demanding higher education to be transformed to meet their changing requirements and demands. Additionally, during Covid-19 pandemic, universities without large endowments, donations, and high tuition found themselves strapped for money. Many rely on government support and subsidies, and with economies growth is shrinking due to Covid-19, governments are pressuring public schools to adopt more sustainable campuses and be more commercially viable. Online education and virtual labs are a much more need economical digital transformation that could differentiate institutions and make them more financially independent [8].

If we learnt anything from this pandemic is that higher education institutions must evolve or become extinct. Adopting digital transformation not only for operational aspects, but for educational pedagogy is a must to survive. Creating online facilities including virtualization, cloud services and producing technology-enhanced curricula that emphasises creativity and workplace skills, will improve educational quality, and reduces costs [9].

In this paper we discuss different technological solutions for providing virtual labs for information security courses. The labs are used in an ethical hacking course at Zayed University, which ran for five semesters. The course includes four to five victim machines that are vulnerable-by-design as well as a final machine used in a gamified project. Students are led through hacking these sample machines in a controlled laboratory setting. During the pandemic lockdown the instructors had to come up with creative solutions for carrying the labs online.

RELATED WORK

The Covid-19 pandemic has forced a new reality on many sectors, especially higher education. Online learning was imposed during the early months of the pandemic and blended classrooms with social distancing are still the norm in most universities today. In blended or hybrid classrooms half of the students would come to campus while the other half study online and interact with their counterparts using teleconferencing technologies; the two groups switch locations every week.

For disciplines such as medicine, dentistry, engineering, and computer science to name a few, they require laboratory sessions, and therefore the challenge of online learning was greater. In those fields the skills depend on the frequent interaction between the professor, students, equipment, components and/or patients. Students access to patients or laboratories is crucial for attaining the experiential learning outcomes, especially in junior and senior years [10].

Virtual labs are a simulation of the physical laboratory environment using computing technologies where students can remotely perform experiments as they would do in a real laboratory. Virtual labs are not new, and numerous attempts have been made in various fields to develop such technology [11]. However, there was a great hesitancy to integrate virtual labs into classes up until the current pandemic lockdowns made them a very important part of e-learning [12]. In many fields such as chemistry and chemical engineering the use of equipment is essential for the students to develop their practical skills. Virtual labs had contributed to development of those same skills, However, without being a complete and effective replacement for the physical labs [13]. Nonetheless, virtual labs have come in handy in engineering, chemistry, and biology classes, because of the flexibility, cost effectiveness and safety [14].

Problem solving is an essential skill for engineers and computer scientists and laboratories play an integral role in developing this skill. Engineering workshops at Massey University, New Zealand had to move online, using Zoom and shared Google Docs. While students were able to work remotely and in teams, the lack of proper virtual lab facilities made it more difficult to replicate the real-life experience [15]. Networking simulations such as NS, OPNET, OMNET, and Cisco Packet Tracer have long been used in teaching computer networking labs and for modeling and testing [16]. In some mechanical engineering disciplines such as manufacturing it is more challenging to develop online interactive simulations. Although learning outcomes were achieved with adequate satisfaction, the problem of developing a more realistic virtual experience remains open [17].

More recently, immersive technologies such as virtual reality (VR), augmented reality (AR) and extended reality (XR) have been successfully used to transition from physical labs or 2D simulations to 3D and more immersive virtual labs with higher interactivity and better visualizations. While the student's feedback was overly positive and virtual labs were similar to their physical counterparts, the students experience was more challenging and less immersive [18]. A similar study on use of VR-based learning, reported positive feedback on familiarity with lab equipment and less effort on equipment setup and malfunction [19]. A study from Swinburne University of Technology compared face-to-face electronics lab class to augmented reality lab using immersive technologies. The gear included "head-mounted AR devices, bi-directional audio and unidirectional mounted video camera, communication channels, and real-time on-line gesturing AR screen". The results report, real-time teamwork, more engagement, ability to experiment with equipment and components, and sufficient interaction with the instructor [20].

Criticism and challenges for virtual labs focused on "no real-life feel and not teaching students about health and safety [21]. However, both changes have been addressed by "augmented reality and multi-sensorial learning;" and proper safety education. While virtual labs will revolutionise education, gaps and challenges remain to be addressed [21].

# VIRTUAL LABORATORY TECHNOLOGIES

There are many kinds of virtual labs, simulation-based tools have been used for a long time for modeling and testing before the actual product implementation. Remote triggered labs on the other hand are a closer representation of face-to-face labs. Virtual labs have many benefits including, syncing theory and practice, higher customization, scalability, more control, ability to repeat procedures, better monitoring of student's progress, accurate digital recording of measurement, cost savings, students' safety, protecting expensive equipment and precious materials from damage and waste, and simulate hard to carry experiments.

Acquiring experimental skills is very important in information security courses. Our earlier work mapped CDIO (Conceive – Design – Implement – Operate) into an ethical hacking course at Zayed University [22]. Carefully crafted vulnerable machines are designed to teach the students different attack vectors and tools [23]. Those machines are deployed in computer laboratories that are isolated for the university production network and the public Internet. The tools used are dangerous, highly malicious and the resulting network traffic must not leak out to public infrastructure. Yet the students must experience hacking in a realistic production-like environment. Finally, project-based learning (PBL) is very important, and the final assessment is a gamified capture the flag (CTF) project [24].

Once Covid-19 lockdown was imposed, we had to promptly switch to online labs mode without compromising the skill building nature of the course. The immediate solution was to provide the students with the virtual machines to setup on their personal computers and laptops. However, this brought with it a myriad of problems that required us to provide many hours of technical support to the students. Some of the issues encountered were, technical difficulties, storage space limitations, hight CPU and memory usage, incompatibility with various virtualisation platforms, erroneous setup and misconfiguration that could jeopardize the security of the students' machines and leak malicious traffic into the public network. The deployment of traditional virtualisation platforms such as VirtualBox and VMware caused the students to be frustrated and their feedback was negative. Therefore, explored the following options over the subsequent semesters since the start of the pandemic [25].

1. *Virtualization at the hardware level*. Hypervisors shown in Figure 1 create a virtualised hardware for each operating system (OS) that is needed to run on the same underlying hardware. Network Address Translation (NAT) helps keep network traffic isolated. The main drawbacks of this approach are performance and security [25].
2. *Virtualization at operating system level*. Which allows *containers* that give different processes, on the same system, the illusion of having different kernels depending on their *namespaces*. Docker *containers* make it very easy for instructors to distribute lightweight machines, and they offer an isolated container network. However, security, GUI and misconfiguration are the main drawbacks [25].
3. *Cloud Computing* (CC) is the perfect solution for virtual labs with automated provisioning, re/instantiation and teardown. However, private clouds must be run at the enterprise level and public clouds would be costly for students or small academic institutions [25].
4. *Vagrant* allows the automation of the configuration of hardware virtualization (type 1) and dockers (type 2) [25].

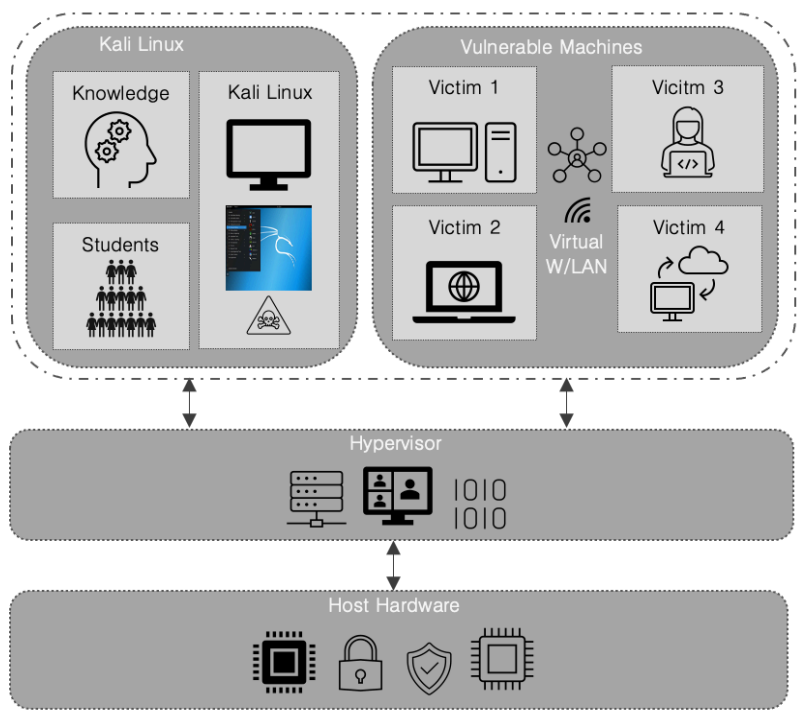

Figure 1: Hardware level virtual lab for ethical hacking course.

In the first two semesters we used hardware virtualization (type 1), in semester 3 we used *dockers* (type 2) and then we transitioned to *Vagrant* (type 4) in semesters 4 and 5. *Could computing* remains under investigation as the private cloud setup is costly, complicated and time consuming, while public clouds offer limited free usage.

RESEARCH METHODOLOGY AND DATA

The subjects of the study were fourth year undergraduate information security major students enrolled in senior level ethical hacking course. The course ran over five semesters in pure online and blended formats, with 15 weeks per semester and two 1.5 hours sessions per week. The progress, opinions and student perceptions were observed, closely monitored, measured, analysed, and the results interpreted. This research followed quantitative methodology and data were gathered by questionnaire on a five-point Likert scale. The survey was validated by several faculty members and graduate students and divided into five sections. The first category of questions covered demographics. The second categorical questions were about the course: syllabus, lab, delivery, and grading. The third category was about teaching and learning outcomes: approach, resources, demos, communication, and writing and technical skills. The fourth category was about the instructor and quality of instruction: time and quality of feedback on assessments, clarity, atmosphere, effectiveness of educational technology and overall quality of instruction. Final category was about enjoyment, attitudes, and behavioural intention towards the course.

FINDINGS

Over five semester 164 respondents completed the survey, and the data was analysed. It is the contentions of the authors that the results cannot be generalized to other than the aforementioned group, however the findings do support a common trend. Because of space limitations, we focus on questions addressing the following aspects.

- Q1. (Setup) The physical or virtual lab environment.
- Q2. (Satisfaction) Satisfaction with the resources provided to support the learning experience.
- Q3. (Curiosity) Understanding of professional practices in the field of study.
- Q4. (Mastery) Have a better understanding of the subject matter.
- Q5. (Attitudes) Overall quality of learning experience.
- Q6. (Intentions) I would recommend this course to others.

Figure 2 shows that scores for the aforementioned questions per semester and includes the consolidated scores over the five semesters. The virtual laboratory setup and value scored 4.26 out of 5, while satisfaction with the resources to support the learning experience was 4.34. The overall score for measuring curiosity and attainment of professional practices was 4.27 and the mastery of the subject matter was 4.34. The score for attitudes towards the overall quality of my learning experience was 4.49 and intentions to recommend this course to others was 4.48.

The total scores are average and have been weighted down by the low scores of the first two semesters. The first half of the first semester was taught face-to-face while the second half we had to abruptly switch to online teaching. The students were shielded from all the technical know-how and halfway through the semester they needed, and were not ready, to assume the full responsibility of setting up their lab environment at home. Students were overwhelmed with technical issues and that increased their frustration. The second semester used hardware virtualization (type 1), which suffered from performance, compatibility, and performance issues. All of the above contributed to lower-than-expected scores for the first two semesters.

The striking result was that as we fully moved towards online, container and vagrant based virtual lab environments, the trend was on the rise with regards to setup, satisfaction, curiosity, mastery, attitudes, and intentions towards the virtual lab course. In Spring semester of 2021, the overall, the virtual laboratory setup was up to 4.63 out of 5, while satisfaction with the resources to support the learning experience was 4.67. Curiosity and attainment of professional practices scored 4.56 and the score for mastery of the subject matter increased to 4.56. The attitudes towards the overall quality of learning experience rose to impressive 4.89 and intentions to recommend this course to others was impressing 4.93. The same scores continued the rise for summer of 2021 as follows: 4.65, 4.67, 4.61, 4.61, 4.94 and 4.97, respectively.

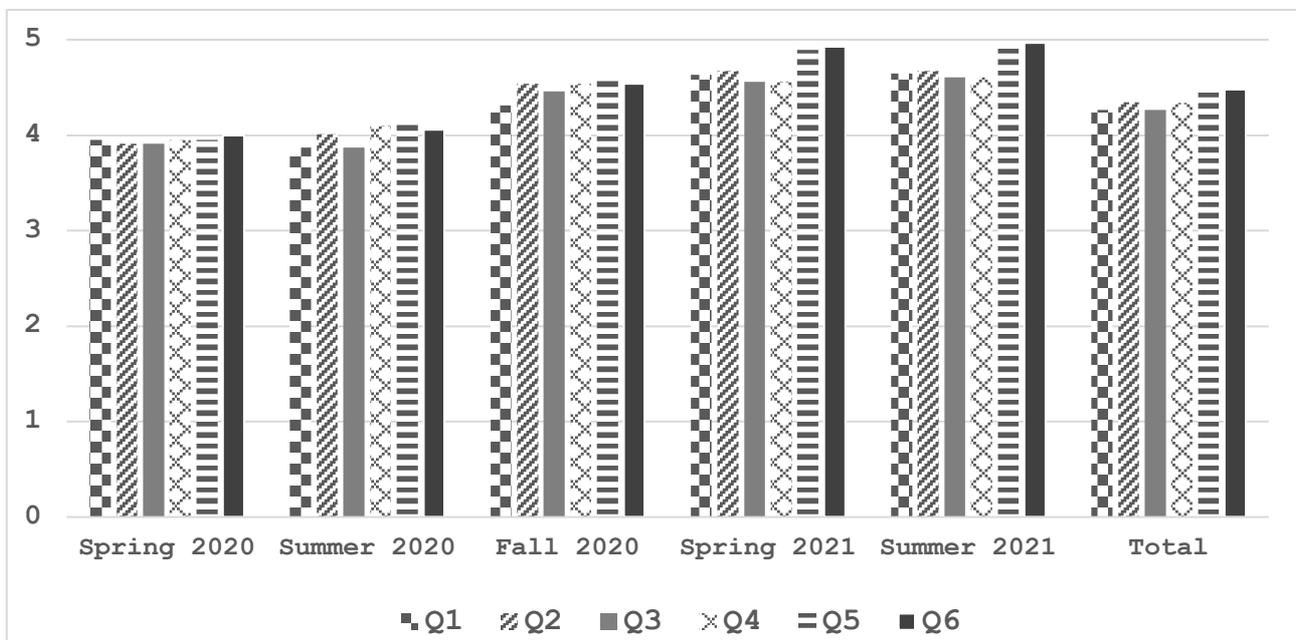

Figure 2: Setup, satisfaction, curiosity, mastery, attitudes, and intentions towards the virtual lab course.

CONCLUSIONS

In conclusion, development of practical experience and technical skills through laboratory are very important in information security courses. The Covid-19 pandemic imposed e-learning and blended classes on most institutions. Virtual labs offered a suitable solution for this predicament. However, the transition from physical face-to-face laboratory experience to virtual labs was not without limitations. Student and instructors were inundated with technical problems, performance issues, connectivity, configurations and time issues. Ethical hacking senior course at Zayed University was offered via several virtual lab technologies. While the feedback was concerning at the beginning of the pandemic the setup, satisfaction, curiosity, mastery, attitudes, and intentions towards the virtual lab course has surpassed expectations. The use of light weight automated technologies has proved to make virtual labs as effective as face-to-face laboratories.

RESEARCH LIMITATION AND FUTURE WORK

Like any survey this study is subject to measurement errors, additionally, convenience sampling bias prevented us from generalising the findings to other populations. However, we believe the trends presented in the finding remain valid and ascertain the efficacy of deploying virtual labs as an alternative for physical labs. Advancements in technology, especially immersive and virtualization technologies call for further investigation. One technology we experimented with was remote access to physical labs machines, which we had to abandon because of security and connectivity issues. Public or private clouds remain as a strong candidate for future research work.

BIOGRAPHY

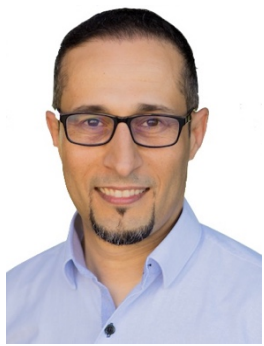

Monther Aldwairi is a full professor at the College of Technological Innovation at Zayed University since fall of 2020. Prior to joining ZU in 2014, he was an associate and assistant professor of computer engineering at JUST since 2007. He served as the vice dean of the Faculty of Computer and Information Technology from 2010 to 2012 and was the assistant dean for student affairs in 2008. In addition, he was an adjunct professor at New York Institute of Technology (NYiT) from 2009 to 2012. Prof. Aldwairi worked at NCSU as post-doctoral research associate in 2007 and as a research assistant from 2001 to 2006. He worked as a system integration engineer for ARAMEX from 1998 to 2000.

Prof. Aldwairi expertise span multiple areas from digital circuit design, computer architecture, information and network security, and software development. Prof. Aldwairi's research interests are in information, network, IoT and web security, intrusion detection, digital forensics, reconfigurable architectures, artificial intelligence, and pattern matching.